\documentclass[journal]{IEEEtran}
\usepackage{cite}
\ifCLASSINFOpdf
\else
\fi
\usepackage{amsmath}
\usepackage{algorithm} 
\usepackage{algpseudocode} 

\usepackage{url}

\usepackage{array}
\ifCLASSOPTIONcompsoc
 \usepackage[caption=false,font=normalsize,labelfont=sf,textfont=sf]{subfig}
\else
 \usepackage[caption=false,font=footnotesize]{subfig}
\fi

\usepackage{stfloats}
\usepackage{url}
\usepackage{graphicx}

\usepackage{amsmath,amssymb}
\usepackage[flushleft]{threeparttable}
\usepackage{array,booktabs,makecell}
\usepackage[font=small]{caption}

\begin{document}
\title{Demonstration of Programmable Brain-Inspired Optoelectronic Neuron in Photonic Spiking Neural Network with Neural Heterogeneity}
%%%% EXAMPLE AUTHORS %%%%
% \author{Michael~Shell,~\IEEEmembership{Member,~IEEE,}
%         John~Doe,~\IEEEmembership{Fellow,~OSA,}
%         and~Jane~Doe,~\IEEEmembership{Life~Fellow,~IEEE}% <-this % stops a space
% \thanks{M. Shell was with the Department
% of Electrical and Computer Engineering, Georgia Institute of Technology, Atlanta,
% GA, 30332 USA e-mail: (see http://www.michaelshell.org/contact.html).}% <-this % stops a space
% \thanks{J. Doe and J. Doe are with Anonymous University.}% <-this % stops a space
% \thanks{Manuscript received April 19, 2005; revised August 26, 2015.}}

\author{Yun-Jhu~Lee,
        Mehmet Berkay~On,
        Luis El Srouji,
        Li Zhang,
        Mahmoud Abdelghany, 
        and~S.J.~Ben~Yoo,~\IEEEmembership{Fellow,~IEEE,~Fellow,~Optica}}%

\maketitle

\begin{abstract}
Photonic Spiking Neural Networks (PSNN) composed of the co-integrated CMOS and photonic elements can offer low loss, low power, highly-parallel, and high-throughput computing for brain-inspired neuromorphic systems. In addition, heterogeneity of neuron dynamics can also bring greater diversity and expressivity to brain-inspired networks, potentially allowing for the implementation of complex functions with fewer neurons. In this paper, we design, fabricate, and experimentally demonstrate an optoelectronic spiking neuron that can simultaneously achieve high programmability for heterogeneous biological neural networks and maintain high-speed computing. We demonstrate that our neuron can be programmed to tune four essential parameters of neuron dynamics under 1GSpike/s input spiking pattern signals. A single neuron circuit can be tuned to output three spiking patterns, including chattering behaviors. The PSNN consisting of the optoelectronic spiking neuron and a Mach-Zehnder interferometer (MZI) mesh synaptic network achieves 89.3\% accuracy on the Iris dataset. Our neuron power consumption is 1.18 pJ/spike output, mainly limited by the power efficiency of the vertical-cavity-lasers, optical coupling efficiency, and the 45 nm CMOS platform used in this experiment, and is predicted to achieve 36.84 fJ/spike output with a 7 nm CMOS platform (e.g. ASAP7) integrated with silicon photonics containing on-chip micron-scale lasers.
\end{abstract}


% \fi
\IEEEpeerreviewmaketitle

\section{Introduction}

Edge artificial intelligence (AI) aims to solve complicated tasks with low latency and under a limited power budget. Hebbian learning processes can take advantage of the heterogeneity of neural dynamics to strengthen and select the most suitable neural pathways for robust learning and improved task performance \cite{PerezNieves2021, GJORGJIEVA201644}.  In fact, the heterogeneity of biological neurons is so vast that a systematic classification scheme has yet to be universally recognized\cite{Zeng2017}. Despite this, all previous photonic and optoelectronic artificial neural network demonstrations have only shown homogeneous (identical) neurons.  Co-integration of CMOS circuitry and optoelectronics can allow heterogeneity and programmability of neurons by adjusting circuit parameters (e.g., bias voltages of various nodes in the neuron circuit).  This paper pursues the design, simulation, and demonstration of heterogeneous co-integrated CMOS-photonic neurons.

\subsection{Heterogeneous Neuron Dynamics}

A regressive perspective on the functional expression of neural networks can be derived from the universal approximation theorem. It states that a neural network composed of a series of neural network layers with any non-affine, continuously differentiable activation function is able to approximate any function with arbitrary accuracy with sufficient neural network width \cite{Hornik1989} or depth \cite{Lu2017}. From this perspective, adapting the set of connection strengths between neural network layers is sufficient to compute any function. Gradient-descent-based methods can be used to solve for the composition of nonlinear weighted sums that best approximate the desired function. This regressive approach, however, encourages a black-box model of design that relies blindly on gradient descent without any guidance for the model to develop structure and functional hierarchy. While working from this framework guarantees that any function can be computed, it offers very little intuition for computing a function with limited time, energy, and computational constraints.

If we instead adopt a functional hierarchy perspective of neural networks, we expect the network to employ a hierarchy of feature selection such that each layer of neurons develops a feature space that detects the presence of a set of features from its input space. Deeper layers will develop a feature space that is desirably more complex than that of the incoming layer, and the network will progressively work towards computing some function. From this perspective, a computing task with a known functional hierarchy can be computed most efficiently by a neural network that imposes the same hierarchy.

For a concrete example of functional hierarchy, it is well-known that subsets of neurons in the human visual cortex are feature-selective, forming "simple" receptive fields of orientation-selective neurons in the primary visual cortex (V1) as well as "complex" receptive fields within the same cortical area\cite{Hubel1962}. The exact function and spatial distribution of specific feature detectors are still debated \cite{Carandini2005}, but the co-location of these features of various complexity represents heterogeneity in the spatial domain of these units. In other words, these neurons identify distinct geometric features from their input space (the visual field). As such, the human visual cortex solves the challenge of visual object recognition by employing a spatial hierarchy starting from simple oriented lines and complex shapes and progressing towards increasingly complex geometric shapes.

While traditional deep neural networks are able to efficiently compute neural networks that impose spatial hierarchies through methods such as convolution, computing across both space and time requires high dimensional tensors which are compute-intensive. Spiking neural networks, however, are dynamical systems whose units can recognize both spatial and temporal features. Thus, for spatio-temporal recognition tasks, like those performed by the human brain, spiking neural networks are likely to provide advantages in computational efficiency when implemented on devices optimized for this computing model. Biological neurons have been shown to exhibit a variety of temporally complex spiking patterns which can be summarized by various models of ordinary differential equations; some models can implement multiple spiking patterns by tuning a small set of flexible parameters \cite{Izhikevich2004}. An efficient computing model for spiking neural networks will allow for these parameters to be tuned according to the spatiotemporal function that is computed by a subset of neurons within the neural network.

The heterogeneity of spiking neural dynamics in a neural network allows for more complex spatiotemporal feature selection that allows hierarchical processing with far fewer units than the traditional neural net. Neural networks with heterogeneous neural dynamics can detect and generate a variety of spiking patterns to implement tasks with complex spatiotemporal dependencies. While the additional complexity of spiking neural networks is not as well studied in deep neural networks, one example shows that the heterogeneity of mixed selectivity neurons can result in more flexible neural networks that solve context-dependent tasks even with simple neuron models. Rigotti et al. \cite{Rigotti2010} model a recurrent neural network whose units are selective to a wide span of random temporal and spatial features and show that the network is capable of switching tasks based on context using far fewer units than traditional models. This example is evidence that the heterogeneity of feature spaces---which is tied to the dynamics of the spiking neuron---gives neural networks a greater expressivity that can be exploited for more efficient neural network computation. In the following demonstration, we show that optoelectronic neurons can efficiently implement heterogeneous neural dynamics within a photonic spiking neural network.

\subsection{Photonic Spiking Neural Networks}
In our previous research \cite{Lee2022}, we discussed the advantage of the photonic spiking neural network (PSNN), which can achieve low loss, low power consumption, high throughput, and high-speed computing. We also presented and demonstrated a monolithic optoelectronic spiking neuron design with both excitatory and inhibitory inputs that shows great promise for future spiking neural networks with great energy efficiency. The previous neuron, however, lacks the tunability to generate multiple patterns of selectivity. As a result, we stated in \cite{9928320} that we simulate a new optoelectronic spiking neuron for the brain-inspired neural network capable of showing different output patterns in the same neuron by tuning several bias voltages. In this paper, we experimentally demonstrated the optoelectronic neuron design in \cite{9928320} using the GlobalFoundries (GF) 45SPCLO process. We present the result of tuning four bias voltages in the neuron circuit and connect it with an MZI mesh synaptic network to show a scalable PSNN with Iris flower classification.

Several research efforts are focusing on the development of spiking neurons. X. Guo \textit{et al.} \cite{10158341} showed cascaded spiking neurons with time domain programming. Another work presented a III-V laser spiking neuron that accepts 25 GBaud input signals \cite{Diamantopoulos:22}. Newns \textit{et al.} \cite{Newns:23} utilized Vertical-cavity surface-emitting laser (VCSEL) dynamics as a spiking neuron unit. W. Zhang \textit{et al.} \cite{10232495} report a spiking neuron composed of two resonant tunneling diode-photodetectors (RTD-PDs) that optically receive and generate spike trains with time-division multiplexed weighting. Some have also explored spiking neurons within photonic neural networks. For example, \cite{Wen:23} exploited the self-pulsing effect in the microring and implemented threshold control to achieve a 19GHz spiking rate PSNN. Meanwhile, \cite{10168617} uses CMOS all-electrical SNN with an excitatory and inhibitory Leaky Integrate-and-Fire (LIF) neuron architecture to achieve 114.90 pJ/inference on the MNIST dataset. However, all the examples above lack tunable neuron parameters, which limits the programmability of the SNN. Rather, research focusing on neuron tunability is rare and has only been demonstrated electrically at low speed---for example \cite{10.1063/5.0151312} employed transistor neurons running at 20Hz. As a result, a suitable neuron for high-throughput computing in PSNN with tunable dynamics has not yet been demonstrated. Our research results show that we can simultaneously achieve high throughput and programmability to provide mixed selectivity neurons for the PSNN.

In the following sections, we present a heterogeneous photonic-electronic neuron design and demonstrate its tunable functionality. Section II introduces the neuron circuit mechanism and its achievable behaviors. Following are the results of programming four tunable circuit nodes with different bias voltages that control different neuronal behaviors. We also present another neuron circuit design extending the previous neuron circuit that can output three different spike patterns. Next, in Section III, we built a PSNN consisting of a 4x4 MZI mesh and a neuron to represent a single-layer neural network. The excitatory and inhibitory functions of the neuron are experimentally verified, and the neural network is applied to the Iris flower classification dataset. We also present a simulation of a potential future neuron that utilizes 7nm technology to further improve energy efficiency. A discussion of the current state-of-the-art heterogeneous neuron is in Section IV.

%%%%%%%%%%%%%%%%%%%%%%%%%%%%%%%%%%%%%%%%%%%%%%%%%%%%%%%%%%%%%%%%%%%%%%%%%%%%%%%%%%%%%%%%%%%%%%%%%%%%%%%%%%%%%%%%%%%%%%%%%%%%%%%%%%%%%%%%%%%%%%%%%%%%%

\section{Brain-inspired Programmable Optoelectronic Neuron Circuit Mechanism and Behavior Tunings}

This section shows the neuron circuit design and its spike generation mechanism. Next, we define a standard neuron input and show the resulting output spike pattern. Then, we show how tuning each control voltage affects the output behavior of the neuron.

\subsection{Neuron circuit firing mechanism}

\begin{figure*}
    \centering
    \includegraphics[width=\linewidth]{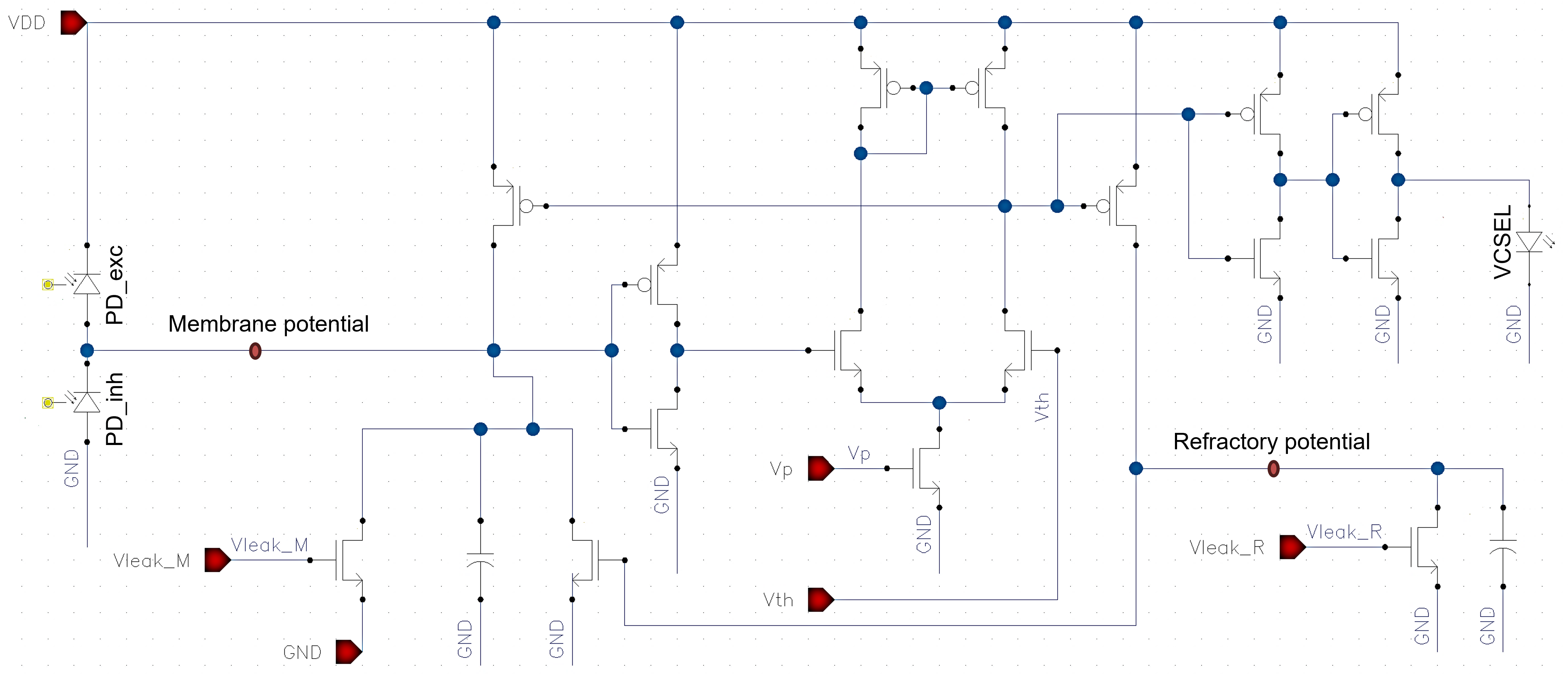}
    \caption{Neuron circuit schematic. The red pins represent the power supply, bias tuning port, and ground to the neuron circuit.}
    \label{fig:neuron_circuit}
\end{figure*}

Our neuron circuit is shown in Fig. \ref{fig:neuron_circuit}. Two photodetectors independently receive optical inputs on the left, representing excitatory (top) and inhibitory (bottom) input. The node between these photodetectors represents the neuron's membrane potential; the voltage of this node determines the overall behavior of the neuron. We have expanded upon the neuronal circuit shown in our previous work \cite{Lee2022,9928320} and experimentally demonstrate an improved design with four tunable parameters. When the membrane potential reaches the threshold determined by $V_{th}$, the neuron generates a spike output and accumulates a refractory potential. The refractory potential will reduce the membrane potential and inhibit the spike output generation for a short period. This competition between membrane potential, refractory potential, and threshold potential describes the basic spike generation mechanism of the neuron circuit. The neuron generates an output electrical signal to drive a VCSEL on the right side of Fig. \ref{fig:neuron_circuit}. Except for VDD and GND, each red pin represents a tunable bias voltage node of the neuron circuit. VDD and GND are the power supply and ground, respectively, while the four bias controls are labeled as follows: $V_{th}$, $V_{leak_M}$, $V_{leak_R}$, and $V_{p}$.

The circuit consists of a positive feedback loop and a negative feedback loop. The positive feedback controls the timing of spike generation and amplifies the output signal in the circuit. The tunable points related to positive feedback are $V_{leak_M}$, $V_{th}$ and $V_{p}$. $V_{leak_M}$ controls the decay rate of membrane potential, which affects the integration window and spike accumulation rate of the neuron. Meanwhile, $V_{th}$ directly controls the neuron threshold at which the neuron will generate a spike. A higher neuron threshold means that a higher membrane potential is needed to initiate a spike output. $V_{p}$ is the positive feedback control and tunes the strength of the positive feedback to the neuron output. On the other hand, the negative feedback loop represents the refractory control of the neuron. The refractory potential delays the ability of the neuron to generate subsequent spikes. $V_{leak_R}$ controls the leakage rate of this refractory potential. Similar to $V_{leak_M}$, it will also affect the accumulation rate of the neuron's refractory potential. 

In combination, these four tunable bias nodes represent the parameters of the neuron's behavior. The resultant programmability of our optoelectronic neuron allows neurons to generate spiking behaviors beyond the leaky integrate-and-fire (LIF) neuron model, which has a simple relationship between input and firing rate and cannot replicate more complex behaviors observed in biological neurons such as chattering \cite{Izhikevich2004}.

%%%%%%%%%%%%%%%%%%%%%%%%%%%%%%%%%%%%%%%%%%%%%%%%%%%%%%%%%%%%%%%%%%%%%%%%%%%%%%%%%%%%%%%%%%%%%%%%%%%%%%%%%%%%%%%%%%%%%%%%%%%%%%%%%%%%%%%%%%%%%%%%%%%%%

\subsection{Neuron circuit behavior}

\begin{figure*}
    \centering
\includegraphics[width=\linewidth]{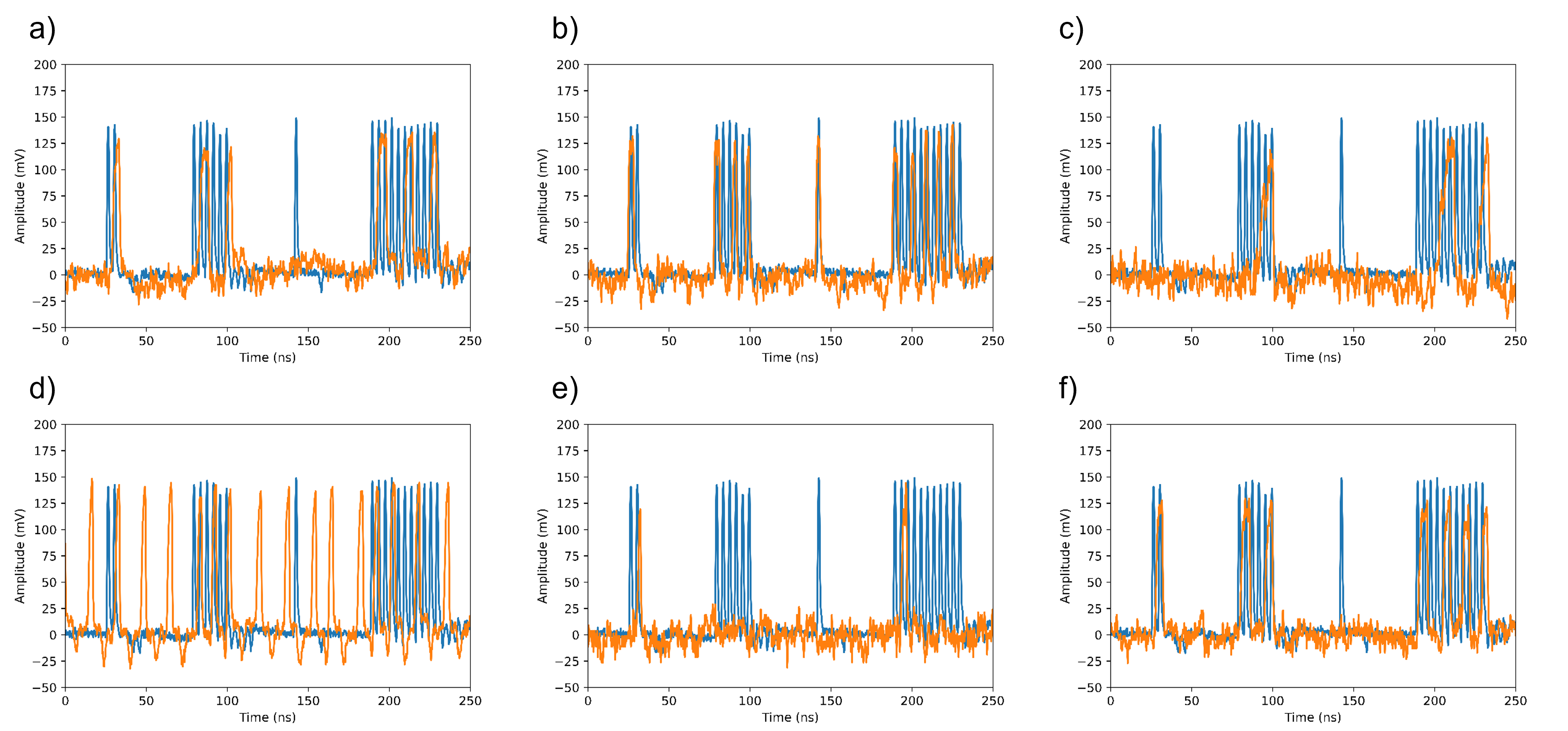}
\caption{a) Standard neuron spiking dynamics with input (blue) and output (orange). The neuron spike dynamics is set as two spike inputs generate one spike output. b) Neuron spiking behavior when threshold biasing is set higher than standard biasing value. The $V_{th}$ biasing in the circuit is designed to be inverted of threshold behavior. c) Neuron spiking behavior when threshold biasing is set lower than standard biasing value. d) Decreased membrane potential leakage biasing will lead to less leakage on membrane potential, which means that the neuron generates spike output faster. e) Decreased refractory potential leakage biasing will lead to higher refractory potential, which means that the neuron generates the next spike output more slowly. f) An increase in positive feedback biasing generates spike output faster.}
\label{fig:n_tuning}
\end{figure*}

Because the heterogeneous spiking neuron can produce a continuum of responses, we define a standard input-output spiking pattern for comparison. We consider the most general case to be a neuron that detects the coincidence of two spikes within a short time window and generates a single output spike in response---Fig. \ref{fig:n_tuning} a) shows this standard coincidence detection pattern. We then explore the effect of each tunable circuit parameter on this standard behavior. The neuron's input comes from a modulated laser source with a peak amplitude of 0.2mW and 1ns spike width at 1GSpike/s spiking rate. This neuron standard pattern has one spike refractory period, as seen from the continuous spikes in Fig. \ref{fig:n_tuning} a).

When designing a full system employing the spiking neuron, the standard parameters of the neuron should be chosen based on the expectations of the neural network architecture. For example, for some systems, it may be desirable for the neuron to respond to stimuli quickly. At the same time, in other cases, it will be preferable for neurons to accumulate more information before a spike is generated. The tunable parameters on our neuron provide the possibility of achieving these different neuron dynamics in a single neuron design. For instance, the tradeoff between the allowed time window of integration for coincide detection and the overall speed of the neural network will need fine-tuning on the neuron dynamics. However, determining the desired neuron dynamics may be difficult because the designer must consider the rate at which inputs are generated, the potential timing jitter of spike trains, the maximum allowed latency for the system output, and more. Below, we provide an example of how to choose the neuron $V_{th}$ parameter.

\begin{figure}[!h]
\centering
\includegraphics[width=\linewidth]{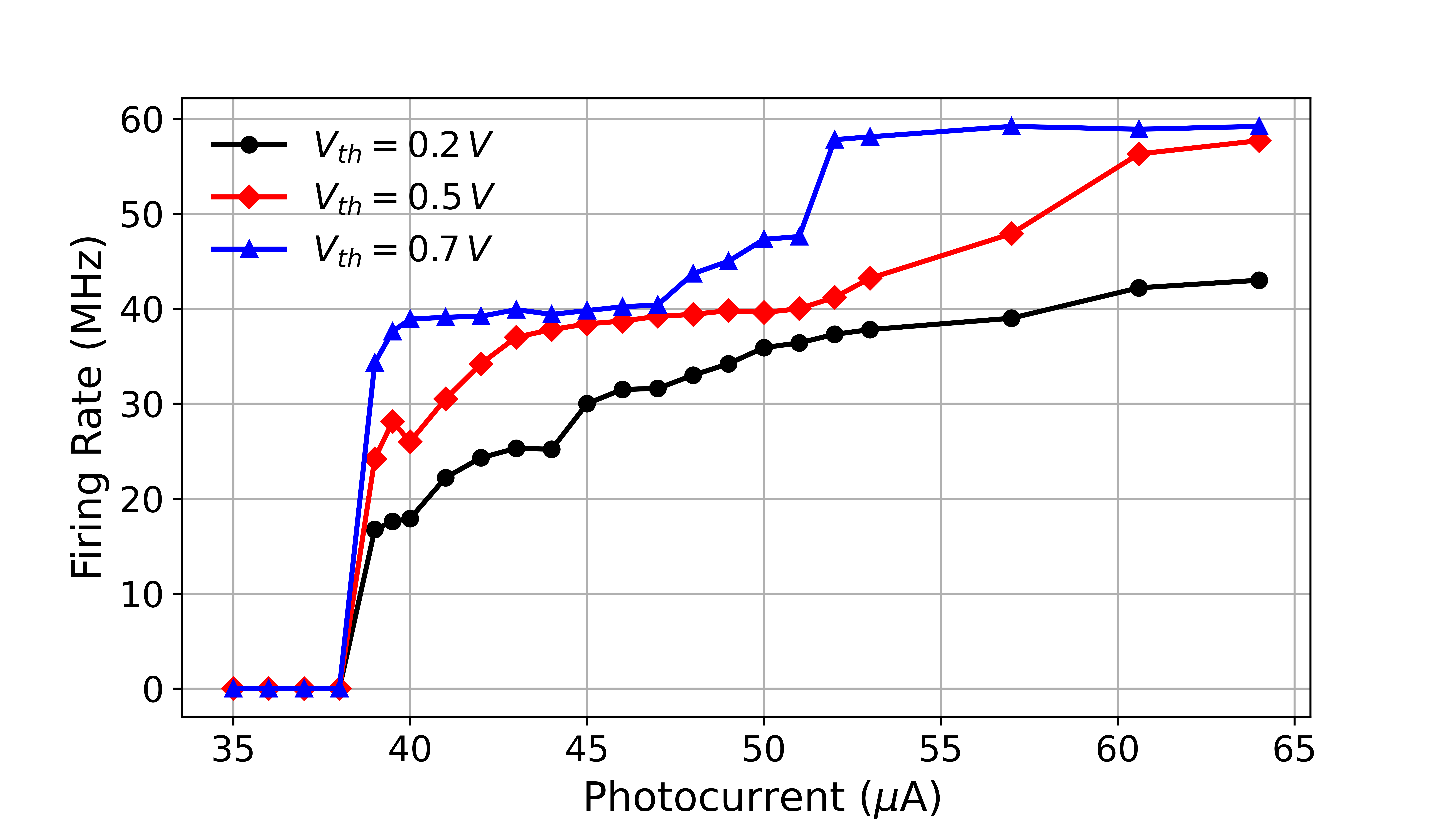}

\caption{Neuron output firing rate corresponding to the excitatory input amplitude at different threshold biasing.}
\label{fig:cw_firing}
\end{figure}

In our case, the $V_{th}$ parameter was chosen according to the qualitative smoothness of the relationship between firing rate and input strength. This smoothness corresponds with a monotonic derivative, which may improve neural network training. We record the neuron output firing rate with the change of input amplitude at $V_{th}$ = 0.2V, 0.5V, and 0.7V. As Fig. \ref{fig:cw_firing} shows, the neuron output firing rate will increase with input amplitude since the membrane potential accumulates faster. However, we can observe that the firing rate increase is not linear for $V_{th}$ = 0.5V and 0.7V. As a result, choosing $V_{th}$ = 0.2V is a better choice for our neuron's standard spiking behavior. 

Please note that this does not mean a NN design should not choose $V_{th}$ = 0.5V and 0.7V in every circumstance. The neuron spiking dynamic should be chosen based on the context of the neural network and its application. The following sections show the tuning results of our neuron's four main tunable parameters. The parameters include threshold tuning ($V_{th}$), membrane potential leakage tuning ($V{leak_M}$), refractory potential leakage tuning ($V{leak_R}$), and positive feedback tunings ($V_{p}$). For reference, the set of biases for the standard coincidence detection pattern is VDD=1V, $V_{th}$=0.2V, $V{leak_M}$=0.5V, $V{leak_R}$=0.5V, and $V_{p}$=0.35V.

%%%%%%%%%%%%%%%%%%%%%%%%%%%%%%%%%%%%%%%%%%%%%%%%%%%%%%%%%%%%%%%%%%%%%%%%%%%%%%%%%%%%%%%%%%%%%%%%%%%%%%%%%%%%%%%%%%%%%%%%%%%%%%%%%%%%%%%%%%%%%%%%%%%%%

\subsection{Threshold tunings}
The neuron threshold controls the membrane potential at which a output spike is initiated. With the same input current over a given period, neurons with lower thresholds generate a spike output more quickly. In contrast, neurons with higher thresholds must accumulate more input current to generate a spike output. Fig. \ref{fig:n_tuning} b) and Fig. \ref{fig:n_tuning} c) show each of these scenarios by tuning only the $V_{th}$ bias voltage and using the standard parameter setting for the remaining biases. The $V_{th}$ bias control is inverted from this logic, meaning that a higher $V_{th}$ bias corresponds to lower threshold neurons. We provide the same input spiking patterns and observe the neuron behavior at different threshold biases. Compared with the standard spiking output pattern in Fig. \ref{fig:n_tuning} a), the higher threshold bias in Fig. \ref{fig:n_tuning} b) generates more spikes (output firing rate is faster) without affecting the refractory (unresponsive) period. In contrast, Fig. \ref{fig:n_tuning} c) shows the opposite scenario with a low threshold bias that requires five input spikes to generate one output spike. This proves that the $V_{th}$ bias tuning matches our expectation.

%%%%%%%%%%%%%%%%%%%%%%%%%%%%%%%%%%%%%%%%%%%%%%%%%%%%%%%%%%%%%%%%%%%%%%%%%%%%%%%%%%%%%%%%%%%%%%%%%%%%%%%%%%%%%%%%%%%%%%%%%%%%%%%%%%%%%%%%%%%%%%%%%%%%%

\subsection{Leakage tunings}
Leakage current is a common feature of ion channels in biological neurons. A leaky channel helps relax the membrane and refractory potential towards their resting equilibria, and heterogeneity of these leakage rates has been shown to improve neural network performance on tasks with rich temporal structure \cite{bouanane2022impact}. The strength of leakage will also determine the speed of the feedback response by inhibiting deviations from the neuron's resting (inactive) state. In our neuron, $V{leak_M}$ controls the speed of positive feedback, and likewise, $V{leak_R}$ determines the negative feedback speed.

We designed the neuron circuit to independently tune membrane and refractory potential leakage by changing the gate voltage of the NMOS transistors to vary the leak conductance on the membrane and refractory potential nodes. Fig. \ref{fig:n_tuning} d) and Fig. \ref{fig:n_tuning} e) show the effect of tuning the leakage of membrane and refractory potential from the standard spiking neuron pattern. The decrease of membrane potential leakage in Fig. \ref{fig:n_tuning} d) will cause the neuron to accumulate membrane potential more quickly, resulting in dense spiking output. In contrast, decreasing refractory potential leakage will hold the refractory potential at high voltage longer, prohibiting the membrane potential from accumulating and leading to a longer waiting time for the output spike to regenerate. Fig. \ref{fig:n_tuning} e) shows the expected result, in which no continuous spike was generated due to decreased refractory potential leakage. If we tune the leakage in real-time, we can implement the higher-order dynamics of various expressions of voltage-gated-ion channels, but this demonstration is out of the scope of this paper.

%%%%%%%%%%%%%%%%%%%%%%%%%%%%%%%%%%%%%%%%%%%%%%%%%%%%%%%%%%%%%%%%%%%%%%%%%%%%%%%%%%%%%%%%%%%%%%%%%%%%%%%%%%%%%%%%%%%%%%%%%%%%%%%%%%%%%%%%%%%%%%%%%%%%%

\subsection{Positive feedback tunings}
Positive feedback is also a crucial parameter in the neuron circuit design, which controls the neuron's ability to amplify the signal inside the neuron circuit. A stronger amplification will increase the strength of the positive feedback mechanism (corresponding to spike generation) over the negative feedback mechanism (corresponding to refractory period). As Fig. \ref{fig:n_tuning} f) shows, increasing the positive feedback biasing will not affect the first spike generation, which would still require two input spikes to generate one output spike because of the unchanged values of $V{leak_M}$ and $V_{th}$. However, the continuous spiking inputs will generate output spikes at a faster rate since the higher $V_{p}$ allows the positive feedback loop to outcompete the negative feedback of the circuit.

%%%%%%%%%%%%%%%%%%%%%%%%%%%%%%%%%%%%%%%%%%%%%%%%%%%%%%%%%%%%%%%%%%%%%%%%%%%%%%%%%%%%%%%%%%%%%%%%%%%%%%%%%%%%%%%%%%%%%%%%%%%%%%%%%%%%%%%%%%%%%%%%%%%%%

\begin{figure*}
    \centering
    \includegraphics[width=\linewidth]{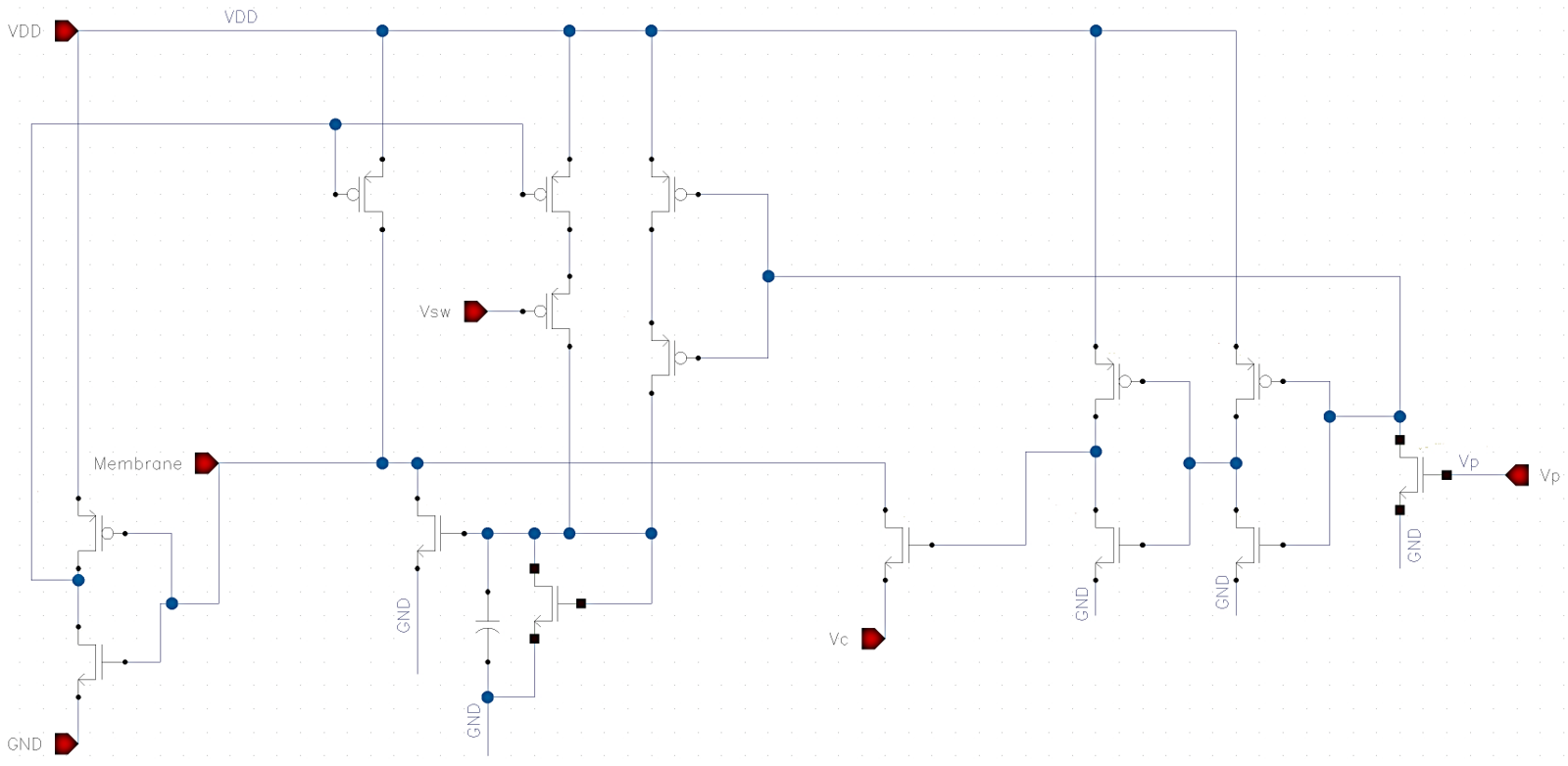}
    \caption{Additional circuit to neuron circuit for different output spiking patterns.}
    \label{fig:neuron_v2}
\end{figure*}

\begin{figure*}
    \centering
    \includegraphics[width=\linewidth]{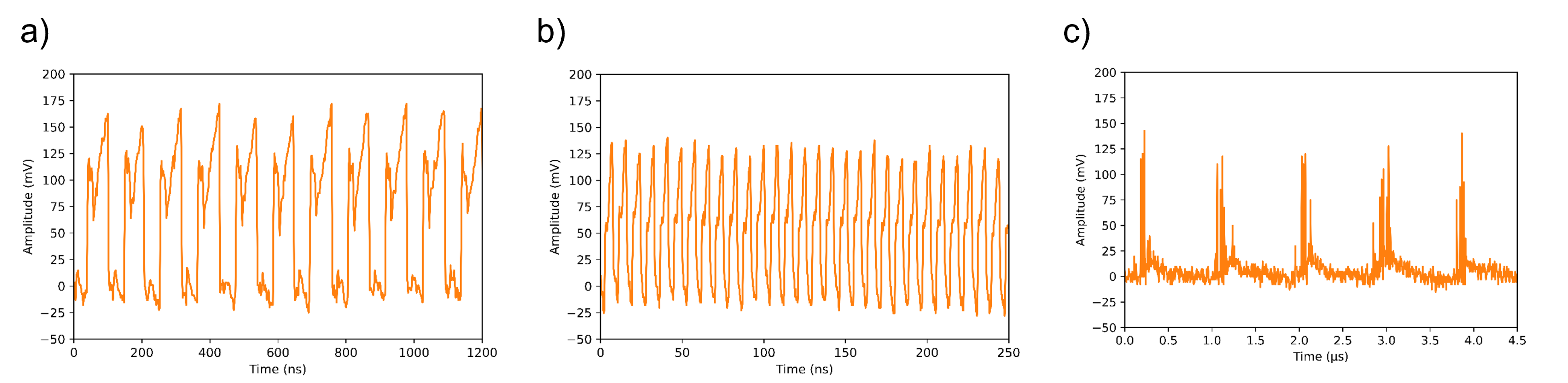}
    \caption{a) Regular spiking  (10MHz spiking output). b) Fast spiking (100MHz spiking output). c) Chattering.}
    \label{fig:neuron_t_v2}
\end{figure*} 

\subsection{Neuron Spiking Output Behavior Tunability}
The above subsection discussed the programmability of the neuron circuit. We show the tuning of threshold, leakage of membrane and refractory potential, and the positive feedback strength. The tuning of spike timing can play a big role in neural network architecture. However, the output spike patterns shown thus far do not fundamentally change. Thus, we also developed another neuron version that outputs different spike patterns to provide greater heterogeneity of neuronal dynamics. This neuron version is based on the neuron discussed in Section II.A and add an additional circuit to tune the output spike patterns. Fig. \ref{fig:neuron_v2} is the schematic of the additional circuit. The red pin "membrane" represents the connection to Fig. \ref{fig:neuron_circuit} membrane potential between two input photodetectors. $V_{p}$ is connected to the positive feedback bias control. $V_{c}$ and $V_{sw}$ are the additional switches to change the neuron output behaviors. $V_{c}$ is the core fine-tuning of membrane potential and $V_{sw}$ is the amplitude switch between negative feedback to membrane.

Fig. \ref{fig:neuron_t_v2} a), b), and c) show three output spiking patterns our neuron generated: regular spiking, fast spiking, and chattering. The neuron under the test receives electrical input from a continuous current source at 0.2mA, unlike the optical inputs in previous sections, while the bias supply tuning $V_{c}$ and $V_{sw}$ differ. We can acquire three output spiking behaviors presented in Fig. \ref{fig:neuron_t_v2}. The regular spiking behavior is similar to the standard spiking behavior in section II and outputs at the rate of 10MHz at the lower end of the firing rate. The fast spiking behavior output at 100MHz with the same step input. The output spiking rate is higher than the standard firing rate in Fig. \ref{fig:n_tuning} a). Additionally, chattering is another output behavior we acquire from this neuron design. As mentioned by Gray and McCormick, “chattering cells may make a substantial intracortical contribution to the generation of synchronous cortical oscillations and thus participate in the recruitment of large populations of cells into synchronously firing assemblies,” \cite{Gray1996}. As a result, the chattering behavior provides an additional mode of communication for spiking neurons and should be included within heterogeneous neural networks. An example set of bias voltages corresponding to each behavior are set as follows:

1)	Regular spiking: threshold ($V_{th}$) low, membrane leakage ($V{leak_M}$) low, refractory leakage ($V{leak_R}$) low, positive feedback ($V_{p}$) low, core fine-tuning ($V_{c}$) low, and amplitude switch ($V_{sw}$) low.

2)	Fast spiking: threshold ($V_{th}$) high, membrane leakage ($V{leak_M}$) high, refractory leakage ($V{leak_R}$) low, positive feedback ($V_{p}$) low, core fine-tuning ($V_{c}$) low, and amplitude switch ($V_{sw}$) high.

3)	Chattering: threshold ($V_{th}$) medium, membrane leakage ($V{leak_M}$) high, refractory leakage ($V{leak_R}$) high, positive feedback ($V_{p}$) medium, core fine-tuning ($V_{c}$) medium, and amplitude switch ($V_{sw}$) high.

%%%%%%%%%%%%%%%%%%%%%%%%%%%%%%%%%%%%%%%%%%%%%%%%%%%%%%%%%%%%%%%%%%%%%%%%%%%%%%%%%%%%%%%%%%%%%%%%%%%%%%%%%%%%%%%%%%%%%%%%%%%%%%%%%%%%%%%%%%%%%%%%%%%%%

\section{PSNN hardware implementation}
Building a spiking neural network requires two primary computational elements: a nonlinear spiking neuron that can integrate its inputs over time, and a synaptic network that can reconfigure to weight connections between these elements. In this section, we connect the optoelectronic programmable neuron with a synaptic network implemented by an MZI mesh previously explored in \cite{9928320}; a single layer of a PSNN was demonstrated to prove the neuron's suitability for this architecture of the optoelectronic neural network. 

\begin{figure}[!h]
\centering
\includegraphics[width=\linewidth]{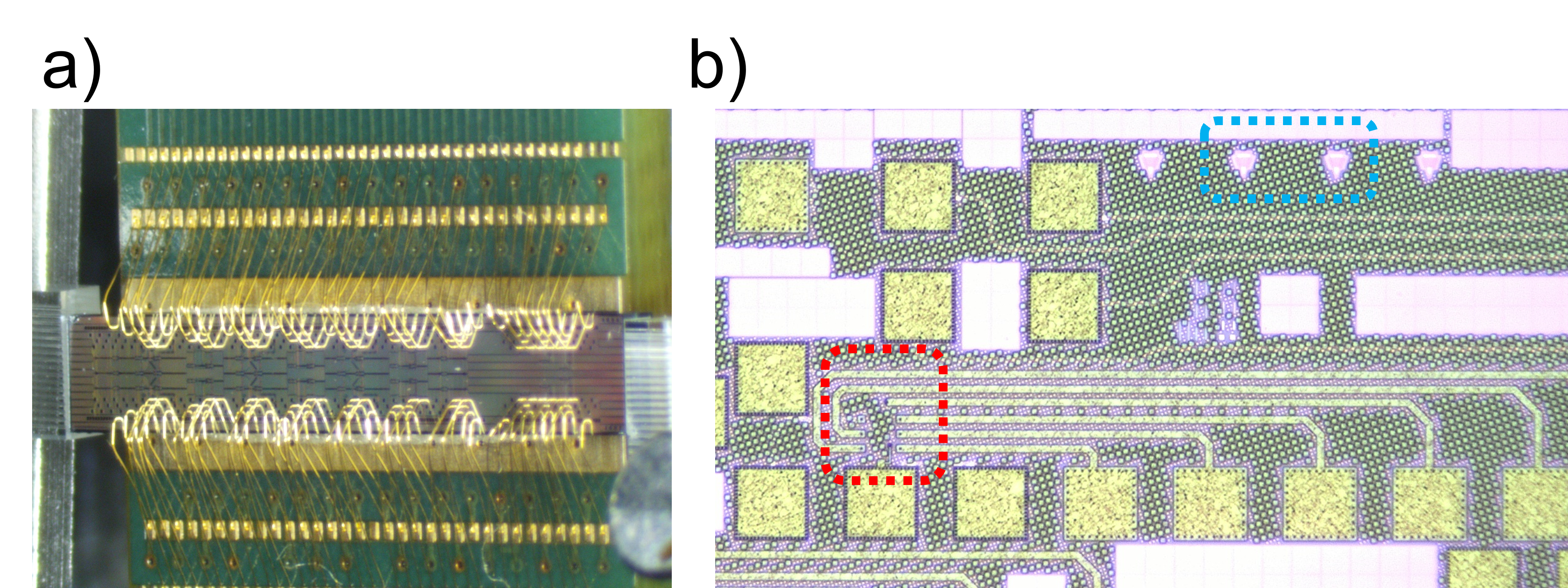}
\caption{a) Photo of MZI mesh fabricated by AIM Photonics. b) Photo of the neuron test structure fabricated by GlobalFoundries 45SPCLO 45nm silicon CMOS photonic process. Optical I/O is labeled in the blue square, and the neuron circuit is labeled in red square.}
\label{fig:pic}
\end{figure}

The photonic integrated circuit (PIC) for MZI mesh and neuron is shown in Fig. \ref{fig:pic}. The neuron was fabricated using the GlobalFoundries 45SPCLO 45nm CMOS process, while the MZI mesh was fabricated by AIM Photonics. The PICs were wirebonded on separate printed circuit boards. MZI synaptic PIC utilizes edge couplers, while the neuron PIC uses vertical couplers. The fiber array alignment setups are set side-by-side on the optical table and connected by fibers. In the future, the PSNN can be monolithically fabricated by the GlobalFoundries 45SPCLO process, which supports silicon photonics and CMOS technologies on a monolithic platform. We use the same standard spiking patterns in Fig. \ref{fig:n_tuning} a) as in the neuron testing in the previous section but use a longer delay between spike groups to avoid interference between spike groups. Subsection A will show the result of excitatory and inhibitory inputs to the neuron, and subsection B will demonstrate offline training on the Iris dataset with multiple spiking inputs to the PSNN simultaneously. 

\subsection{PSNN with neuron's excitatory and inhibitory inputs}
\begin{figure}[!h]
\centering
\includegraphics[width=\linewidth]{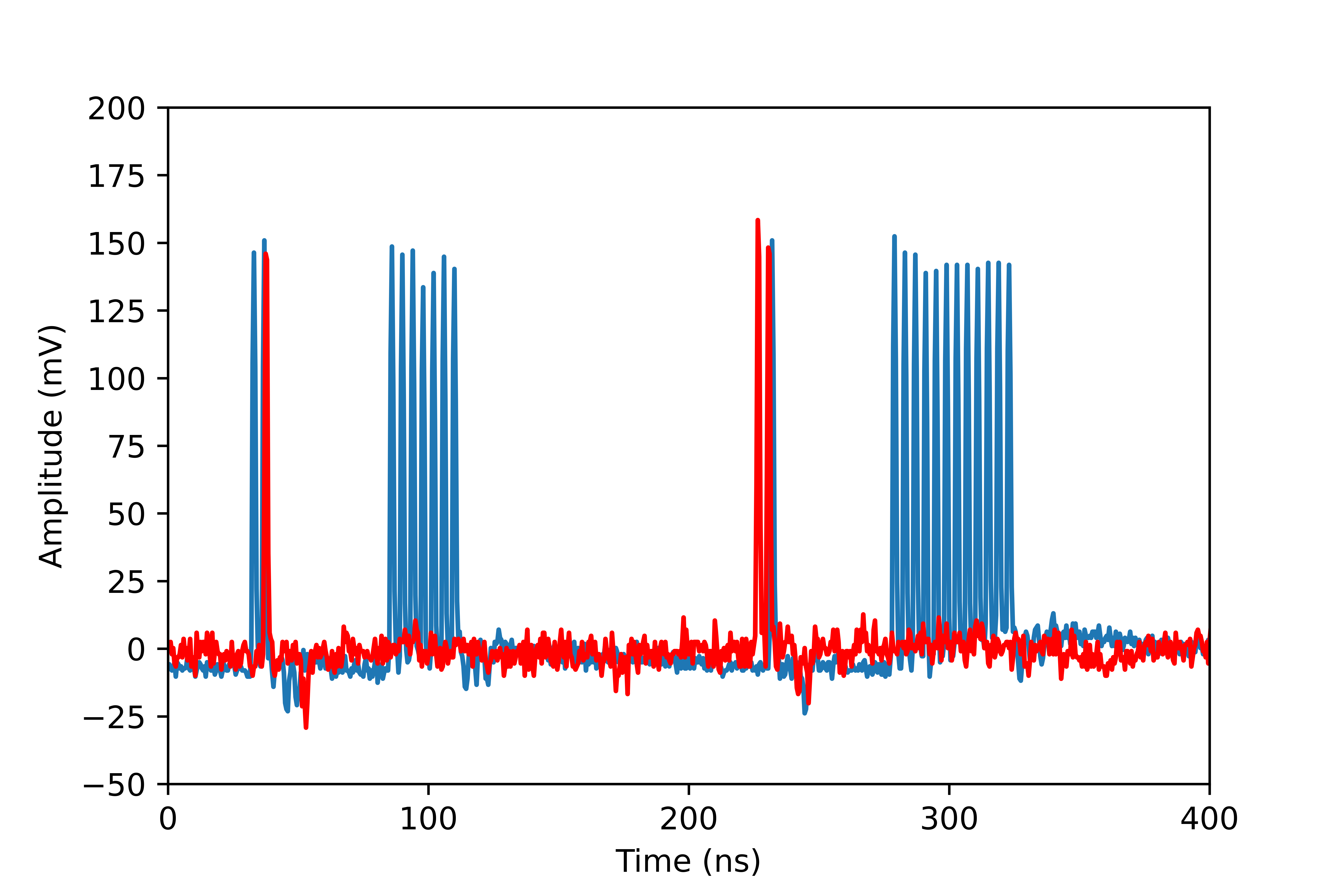}
\caption{Excitatory input (blue) and inhibitory input (red).}
\label{fig:exc_inh}
\end{figure}

\begin{figure*}
    \centering
    \includegraphics[width=\linewidth]{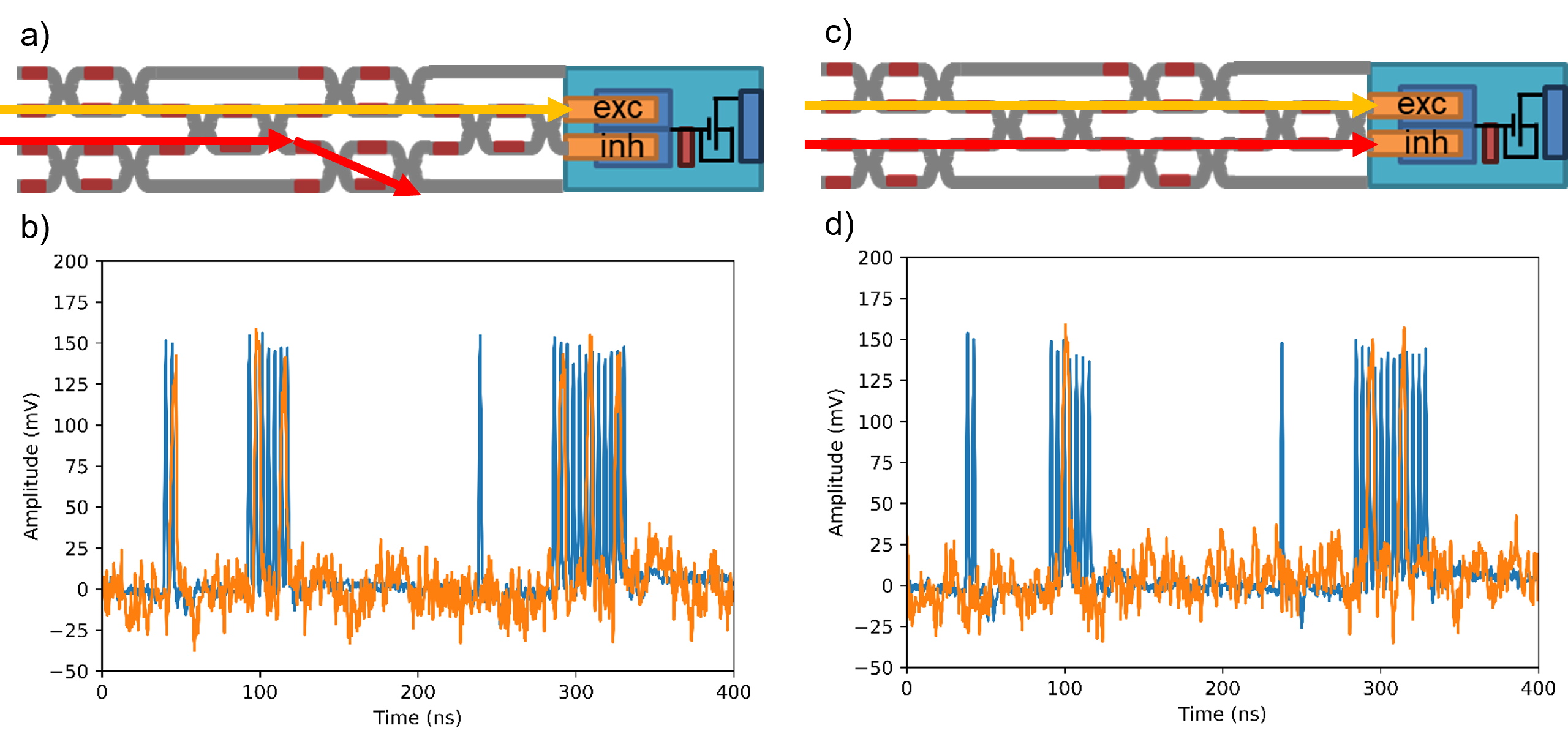}
    \caption{a) Inhibitory laser power drop inside the MZI mesh. b) neuron spiking behavior with inhibitory laser power drop inside the MZI mesh (neuron input (blue) and neuron output (orange)).c) Inhibitory laser all power to neuron inhibitory input. d) neuron spiking behavior with Inhibitory laser all power to the neuron (neuron input (blue) and neuron output (orange)).}
    \label{fig:neuron_e_i}
\end{figure*}

We generate two spiking patterns on two independent lasers to serve as excitatory input and inhibitory input separately. As Fig. \ref{fig:exc_inh} shows, excitatory input remains the same pattern as we used to test neurons in the previous section, and an additional inhibitory input signal is applied to the neural network. Fig. \ref{fig:neuron_e_i} a) and c) are the schematic diagrams showing the routes taken by the spiking excitatory and inhibitory laser inputs. The excitatory route is set the same for both cases, and only the inhibitory route differs. In Fig. \ref{fig:neuron_e_i} a), the inhibitory input power is detoured to a dummy port on the MZI mesh when we set the MZI to the cross-state. In this case, the neuron only receives signals from excitatory input, and as Fig. \ref{fig:neuron_e_i} b) shows, the neuron input/output behavior is similar to the standard spiking pattern in Fig. \ref{fig:n_tuning} a). In the next scenario, Fig. \ref{fig:neuron_e_i} c), the inhibitory input from MZI mesh goes into the inhibitory detector in the neuron, and we observe the resultant change in output spike timing in Fig. \ref{fig:neuron_e_i} d). When the same amplitude of inhibitory and excitatory inputs is presented, the neuron cannot generate a spike output, proving the neuron's inhibitory functionality. Additionally, we see in Fig. \ref{fig:neuron_e_i} d) that the output spike timing on longer input spike trains also changes compared to Fig. \ref{fig:neuron_e_i} b), which is caused by the delay effect of the inhibitory signal on output spike generation.

%%%%%%%%%%%%%%%%%%%%%%%%%%%%%%%%%%%%%%%%%%%%%%%%%%%%%%%%%%%%%%%%%%%%%%%%%%%%%%%%%%%%%%%%%%%%%%%%%%%%%%%%%%%%%%%%%%%%%%%%%%%%%%%%%%%%%%%%%%%%%%%%%%%%%

\subsection{Iris dataset implementation on PSNN}
\begin{figure*}
    \centering
    \includegraphics[width=16.8cm]{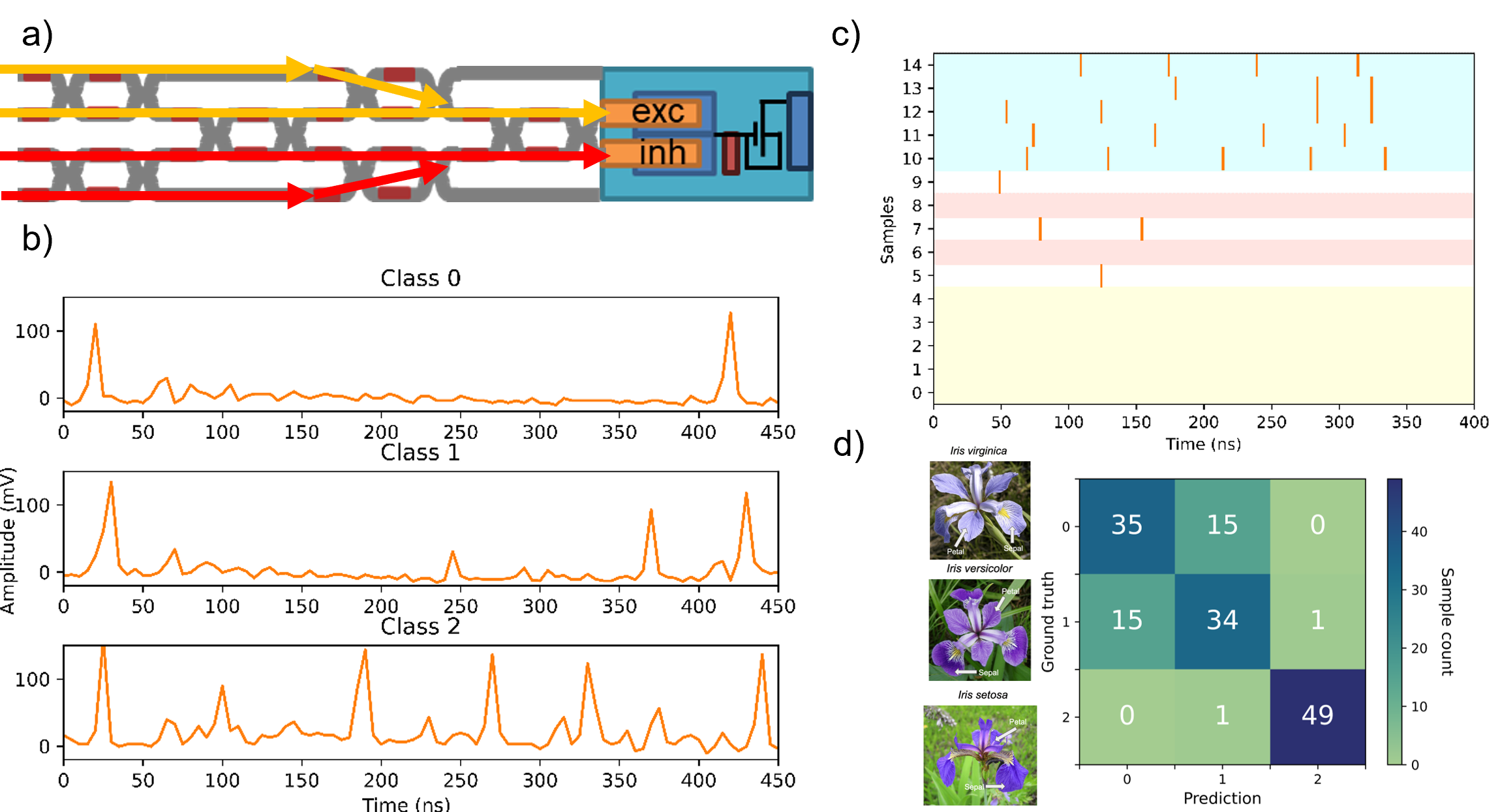}
    \caption{a) 4 inputs with different spiking patterns representing iris sample pump to the neural network simultaneously. b) Iris dataset classification output. Excluding the header at the front (0-50 ns) and the header of the next sequence (400-450 ns), there are zero spikes in class 0 (Iris virginica), less than two spikes in class 1 (Iris versicolor), and more than two spikes in class 2 (Iris setosa). c) Spike time raster plot. Sample 0-4 is class 0 sample inputs, sample 5-9 is class 1 sample inputs, and sample 10-14 is class 2 sample inputs. Samples 6 and 8 were misclassified to class 0 from our definition. d) Iris dataset classification output confusion matrix.}
    \label{fig:iris}
\end{figure*}

To further demonstrate the usage of MZI mesh synaptic network and neuron, we apply our PSNN setup to the Iris flower dataset for classification, consisting of four features and 150 samples \cite{iris}. The four features of each sample are presented as four optically encoded spiking inputs. These inputs simultaneously pump into the neural network to represent Iris dataset samples adjacently in time. Each input to the network is composed of spike trains of 1ns pulse width and varying frequency based on the original Iris dataset feature value. We also pre-trained the weight value for the MZI mesh and found two positive weight values and two negative weight values for the Iris dataset classification. Thus, we assigned the positive weights to excitatory input and negative weights to inhibitory inputs for one Iris sample as Fig. \ref{fig:iris} a) shows.

For the output neuron encoding, we categorized the neuron output spiking number in the 400ns period into three categories. We defined an output of zero spikes as class 0 (Iris virginica), an output of less than two spikes as class 1 (Iris versicolor), and an output of more than two spikes as class 2 (Iris setosa). Fig. \ref{fig:iris} b) shows the output neuron recording for one sample of each class in the Iris dataset. We can read the output spike response accordingly by excluding the header spike (indicating the start of the 400ns sample period) at the front and back (header for the following sample). A spike threshold of 75mV was used to exclude random noise from the spike count.

We achieved 89.3\% accuracy for this offline training method. Fig. \ref{fig:iris} c) shows output spiking results for 15 randomly selected Iris samples, and Fig. \ref{fig:iris} d) is the confusion matrix for the Iris flower classification task. The separation for class 0 and class 1 did not perform well, while we obtained higher accuracy performance for separating class 1 and class 2. The result matches the original dataset in that class 2 (Iris setosa) was shown to be easier to classify, while the separation of class 0 (Iris virginica) and class 1 (Iris versicolor) required a deeper neural network architecture because these classes are not linearly separable on the input space.

%%%%%%%%%%%%%%%%%%%%%%%%%%%%%%%%%%%%%%%%%%%%%%%%%%%%%%%%%%%%%%%%%%%%%%%%%%%%%%%%%%%%%%%%%%%%%%%%%%%%%%%%%%%%%%%%%%%%%%%%%%%%%%%%%%%%%%%%%%%%%%%%%%%%%

\subsection{Energy consumption between Foundry-enabled neuron and futuristic ASAP7 PDK neuron}
To understand the energy consumption of our tunable optoelectronic neuron and make predictions for future designs, we analyze the neuron's current-voltage (I-V) relationship through the Cadence Spectre simulator. The simulator lets us record the real-time power value according to the circuit's I-V response. As a result, we can predict the energy consumption for our tunable optoelectronic neuron by calculating the energy required for each spike output. 

The photodetector for the neuron has 1A/W sensitivity in the simulation. The standard neuron with four tunable parameters requires 1.18 pJ/spike to drive the vertical-cavity lasers with a fanout of 15---corresponding to a peak output laser power of 3mW compared to the 0.2mW peak input power in Section II. A spiking scenario. The fanout can be further increased if the neurons are input with low peak power but higher frequency signals. The revised chattering neuron version consumes 12.84 pJ/spike output for the regular spiking pattern, 4.28 pJ/spike output for the fast-spiking pattern, and 2.38 pJ/grouping of chattering spikes. The above results are based on a 45nm technology node and can be further improved using more advanced technologies. To support this claim, we simulate our neuron circuit using the ASAP7 PDK \cite{CLARK2016105}. ASAP7 contains SPICE-compatible FinFET device models (BSIM-CMG) for the 7nm node. The simulated ASAP7 neuron's spike input and output response result is shown in Appendix Fig. \ref{fig:asap7}, and the comparison for the foundry-enabled neuron and ASAP7 neuron is in Table \ref{table:power}. Each output spike consumes 36.84 fJ/spike.

\begin{table}[!h]
  \caption{Foundry neuron and ASAP7 neuron comparison}\label{table:power}
  \centering 
  \begin{threeparttable}
    \begin{tabular}{ccc}
    %{m{15mm} m{70mm} m{18mm}}
    Types  & Foundry neuron & ASAP7 neuron\\
     \midrule\midrule
technology & 45nm & 7nm  \\
    \cmidrule(l  r ){1-3}
Maximum spiking rate  &   1GSpike/s        &   5GSpike/s    \\%new row
    \cmidrule(l  r ){1-3}
spike width & 1ns & 0.2ns \\
    \cmidrule(l  r ){1-3}
input capacitor &  901fF & 0.128fF  \\
    \cmidrule(l  r ){1-3}
energy/spike output & 1.18pJ & 36.84fJ \\
    \midrule\midrule
    \end{tabular}
%    \begin{tablenotes}
%\item[*]. 
%\end{tablenotes}

\end{threeparttable}
  \end{table}

%%%%%%%%%%%%%%%%%%%%%%%%%%%%%%%%%%%%%%%%%%%%%%%%%%%%%%%%%%%%%%%%%%%%%%%%%%%%%%%%%%%%%%%%%%%%%%%%%%%%%%%%%%%%%%%%%%%%%%%%%%%%%%%%%%%%%%%%%%%%%%%%%%%%%

\section{Tunable neuron heterogeneity}

In this paper, we demonstrate a highly tunable analog optoelectronic neuron capable of operating at 1GSpike/s input spiking rate with an energy efficiency of 1.18pJ/spike. We also use a SPICE model included in the ASAP7 PDK \cite{CLARK2016105} to project a 5GSpike/s spiking rate and 36.84fJ/spike efficiency using a more advanced CMOS process node. With this demonstration, we show that an analog optoelectronic spiking neuron can operate at high speeds comparable to completely optical neuron models while achieving the complexity of spiking neuron models governed by dynamical systems. These models are capable of more complex spatiotemporal feature detection than traditional neuron models.

While many other demonstrations of optical spiking neurons have been previously reported \cite{10158341,Diamantopoulos:22,Newns:23,10232495,Wen:23,10168617,10.1063/5.0151312,Kuszelewicz2011,Hurtado2012,Nahmias2013,Shastri2016,Chakraborty2018,Xiang2019,Peng2020,Huang2022}, none of these works demonstrate spiking behaviors beyond the LIF or resonate-and-fire (RIF) neuron models. Further, many of these models rely on a bijection between the dynamics of a laser or modulator and the dynamics of a neuron model. As a result, these devices are much more constrained to material parameters that are not programmable and, therefore, restrict their use in a heterogeneous neural network. In particular, these lasers require constantly driving laser current while our neurons can achieve near-zero power consumption when the neurons are at rest.

If we instead compare it to the state-of-the-art in programmable neuromorphic computing, this neuron model is faster, more expressive, and more energy-efficient than the Loihi neuromorphic processor \cite{Davies2018}. The Loihi processor adopts an asynchronous many-core digital approach to spiking neural networks. This means that a single spike message will incur multiple energetic costs as it travels across the network-on-chip. Meanwhile, optical networks can achieve high-bandwidth, low-loss transmission networks across great distances. Even restricting the network to a single digital core, Loihi shows an energetic cost of 1.7pJ/spike and a processing capability of only 3.44 Gspikes/s. This means that even at the 45nm process node (compared to Loihi's 14nm node), the optoelectronic approach can process more spikes---scaling as 1GHz*(number of neurons)---at a lower energetic cost per spike. This result demonstrates the feasibility of future optoelectronic neural networks to provide high-speed, programmable deep neuromorphic neural networks.

%%%%%%%%%%%%%%%%%%%%%%%%%%%%%%%%%%%%%%%%%%%%%%%%%%%%%%%%%%%%%%%%%%%%%%%%%%%%%%%%%%%%%%%%%%%%%%%%%%%%%%%%%%%%%%%%%%%%%%%%%%%%%%%%%%%%%%%%%%%%%%%%%%%%%

\section{Conclusion}
We present the design, fabrication, and experimental demonstration of our optoelectronic programmable neuron. Our neuron can process a 1GSpike/s input spiking rate and is equipped with four bias controls that can flexibly tune the neural complexity and change output spiking patterns to provide heterogeneous neural dynamics. We also show a PSNN employing the programmable neuron and achieving 89.3\% classification accuracy on the Iris flower dataset. The neuron based on 45nm foundry technology has an energy consumption of 1.18pJ/spike output and can further improve the performance to 36.84fJ with ASAP7 7nm technology. The neuron can satisfy the need for high-throughput computing and heterogeneity to solve complex tasks on edge neuromorphic devices.

%%%%%%%%%%%%%%%%%%%%%%%%%%%%%%%%%%%%%%%%%%%%%%%%%%%%%%%%%%%%%%%%%%%%%%%%%%%%%%%%%%%%%%%%%%%%%%%%%%%%%%%%%%%%%%%%%%%%%%%%%%%%%%%%%%%%%%%%%%%%%%%%%%%%%
\section*{Acknowledgment}

This material is based upon work supported by the Air Force Office of Scientific Research under award number FA9550-18-1-0186 and award number FA 9550-22-1-0532, and in part by the Office of the Director of National Intelligence, Intelligence Advanced Research Projects Activity under Grant 2021-21090200004.
\setcounter{figure}{0}

%\clearpage

\appendices
\section{ASAP7 simulation results}

\begin{figure}[!h]
\centering
\includegraphics[width=\linewidth]{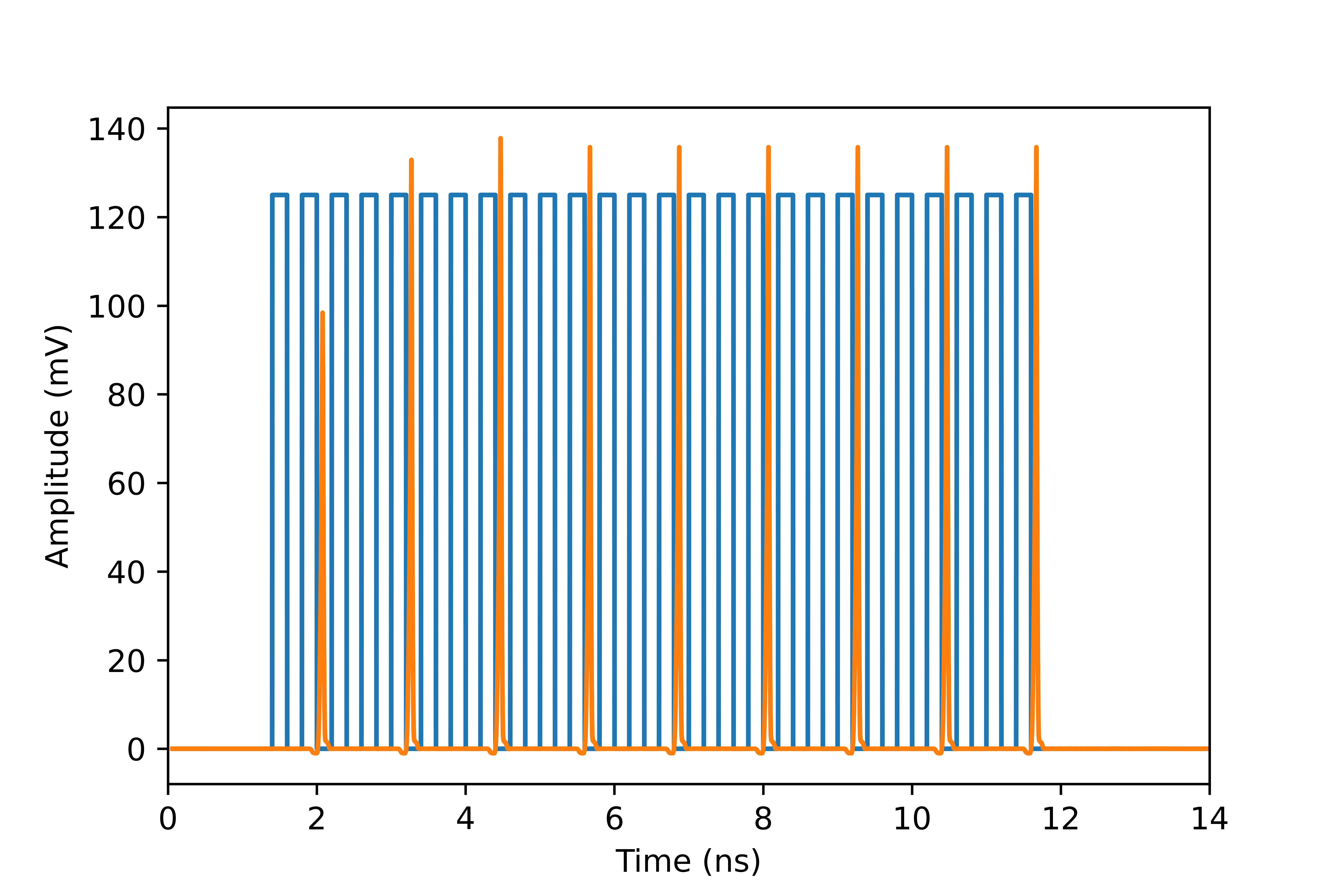}
\caption{ASAP7 neuron simulation of standard patterns.}
\label{fig:asap7}
\end{figure} 

\bibliographystyle{IEEEtran}
\bibliography{Heterogeneity_neuron.bib}



% \end{IEEEbiography}

\begin{IEEEbiographynophoto}{Yun-Jhu Lee}
received the B.S. in Life Science from the National Taiwan University, Taiwan. He is currently working towards the Ph.D degree in Electrical and Computer Engineering at the University of California, Davis. Research interests include neuromorphic computing, integrated photonics, MEMS, and control systems.
\end{IEEEbiographynophoto}

\begin{IEEEbiographynophoto}{Mehmet Berkay On}
received the B.S. in Electrical and Electronics Engineering from the Bilkent University, Ankara, Turkey in 2018. He is currently working towards the Ph.D degree in Electrical and Computer Engineering at the University of California, Davis. Research interests are enerrgy-efficient photonic neuromorphic systems, RF-photonic signal processing, fiber-optic communication, and compressive sensing.
\end{IEEEbiographynophoto}

%\newpage

\begin{IEEEbiographynophoto}{Luis El Srouji}
received the B.S. in Applied Physics with an emphasis in Physical Electronics from the University of California, Davis in 2020. He is currently working towards the Ph.D degree in Electrical Engineering at the University of California, Davis. Research interests include the design of bio-physically accurate analog neuron circuits, development of learning algorithms for optoelectronic spiking neural networks, and fabrication of on-chip laser sources.
\end{IEEEbiographynophoto}

\begin{IEEEbiographynophoto}{Li Zhang}
Li Zhang received the B.S. degree in electronics and information technology and instrumentation from Zhejiang University, Hangzhou, China, in 2016. He is currently pursuing the Ph.D. degree in electrical engineering with the University of California at Davis, Davis, CA, USA. His research interests include ultra-wideband transceiver, trans-impedance amplifier and optical driver.
\end{IEEEbiographynophoto}

\begin{IEEEbiographynophoto}{S. J. Ben Yoo}
(Fellow, IEEE and Fellow, Optica) received the B.S. degree in electrical engineering with distinction, the M.S. degree in electrical engineering, and the Ph.D. degree in electrical engineering with a minor in physics, from Stanford University, Stanford, CA, USA, in 1984, 1986, and 1991, respectively. He is currently a Distinguished Professor of electrical engineering with UC Davis, Davis, CA, USA. His research with UC Davis includes 2D/3D photonic integration for future computing, communication, imaging, and navigation systems, micro/nano systems integration, and the future Internet. Prior to joining UC Davis in 1999, he was a Senior Research Scientist with Bellcore, leading technical efforts in integrated photonics, optical networking, and systems integration. His research activities with Bellcore included the next-generation internet, reconfigurable multiwavelength optical networks (MONET), wavelength interchanging cross connects, wavelength converters, vertical-cavity lasers, and high-speed modulators. He led the MONET testbed experimentation efforts, and participated in ATD/MONET systems integration and a number of standardization activities. Prior to joining Bellcore in 1991, he conducted research on nonlinear optical processes in quantum wells, a four-wave-mixing study of relaxation mechanisms in dye molecules, and ultrafast diffusion-driven photodetectors with Stanford University. He is a fellow of OSA, NIAC, and was the recipient of the DARPA Award for Sustained Excellence (1997), the Bellcore CEO Award (1998), the Mid-Career Research Faculty Award (2004 UC Davis), and the Senior Research Faculty Award (2011 UC Davis).
\end{IEEEbiographynophoto}


%\vfill


\end{document}